\documentclass[12pt,a4paper]{article}
\usepackage{amsmath,amsfonts,epsfig}
\usepackage{amssymb,bbold}
\usepackage{mathrsfs}
\usepackage{hyperref}
\newcommand{\dlk}{ {\tilde{d l}}}
\newcommand{\eps}{\epsilon}

\newcommand{\om}{{\omega}}

\newcommand{\intinf}{\mathop{\int\limits_{-\infty}^\infty}}

\begin{document}
\numberwithin{equation}{section}
\setlength{\unitlength}{.8mm}

\begin{titlepage} 
\vspace*{0.5cm}
\begin{center}
{\Large\bf NLIE formulations for the generalized Gibbs ensemble in the sine-Gordon model}
\end{center}
\vspace{1.5cm}
\begin{center}
{\large \'Arp\'ad Heged\H us}
\end{center}
\bigskip

\vspace{0.1cm}

\begin{center}
HUN-REN Wigner Research Centre for Physics,\\
H-1525 Budapest 114, P.O.B. 49, Hungary\\ 
\end{center}
\vspace{1.5cm}
\begin{abstract}

\end{abstract}
In this paper we propose two sets of nonlinear integral equations (NLIE)  for describing 
the thermodynamics in the sine-Gordon model, when higher Lorentz spin conserved charges 
are also coupled to the Gibbs ensemble. We call them NLIE I and II.

The derivation of the equations, is based on $T-Q$ relations given by the equivalent thermodynamic Bethe ansatz 
(TBA) formulation of the problem in the repulsive regime. Though the equations are derived in the repulsive 
regime at discrete values of the coupling constant, a straightforward analytical continuation ensures their validity 
within the whole repulsive regime of the theory.
For the NLIE I formulation,  
 appropriate analytical continuation makes the penetration into the attractive regime also possible. 
However, the magnitude of this penetration 
is restricted by the spin of the largest spin conserved charge contained in the Gibbs ensemble. 

Within their range of validity, these NLIE formulations provide efficient theoretical frameworks for 
computing expectation values of conserved charge densities, their associated currents, and  
vertex operators and their descendants, with respect to the generalized Gibbs ensemble.

\end{titlepage}

\section{Introduction}

The sine-Gordon model is a paradigmatic example of a $1+1$ dimensional relativistic quantum field theory (QFT), and it has played an important  role in the development of quantum integrable models. It exhibits rich and highly non-trivial dynamics, while its integrability enables the use of powerful non-perturbative techniques \cite{zamzam79,smirnov92,LukZam96} to solve the model. 

Beyond its theoretical appeal, the sine-Gordon model also has experimental relevance, as it can be realized in systems of cold atoms 
\cite{Gritsev:2007}--\cite{expsg5}. 
Further realizations via quantum circuits \cite{expcirc} and coupled spin chains \cite{explatt} are also proposed. 

In recent years, there has been growing interest in the non-equilibrium dynamics of isolated quantum systems, driven by both theoretical progress and experimental advances in cold atom physics. In integrable systems, where an infinite set of conserved quantities constrain the dynamics, conventional thermalization is replaced by relaxation towards a Generalized Gibbs Ensemble (GGE) \cite{Eisert:2014jea,Rigol:2007, Vidmar:2016}. 
A crucial aspect of the generalized Gibbs ensemble is that it provides the fundamental statistical description of integrable systems' long time steady states by accounting for the complete set (or a large subset) of conserved quantities \cite{Rigol:2007, Vidmar:2016, Essler:2016, Ilievski:2016}. 
GGE is not only essential for understanding equilibration in isolated quantum systems but also serves as the foundational building block for advanced theoretical frameworks such as generalized hydrodynamics 
\cite{GHD1,GHD2}. By precisely capturing the local conserved charges and their associated potentials{\footnote{Later, in the text we also call them generalized inverse temperatures in order to avoid confusion with the thermodynamic potentials.}}, the GGE establishes the baseline ensemble from which hydrodynamic equations describing non-equilibrium transport and large-scale dynamics in integrable models are derived. Thus, the significance of the GGE extends beyond static properties, underpinning much of the modern progress in describing integrable quantum dynamics.

Recent research has seen a resurgence of interest in the equilibrium and non-equilibrium thermodynamics of the sine-Gordon model, 
which has become an actively explored domain in integrable quantum field theories 
\cite{Nagy:2023wkz}--\cite{Bastianello:2023}. 
Despite the model's integrability, the formulation of the generalized Gibbs ensemble remains non-trivial. 
While the Thermodynamic Bethe Ansatz (TBA) \cite{Zamolodchikov:1989cf,Caux} provides a framework for constructing the GGE in the sine-Gordon model, 
the structure and form of the resulting equations depend strongly on the value of the coupling constant, making the implementation cumbersome and highly case-specific \cite{FW2, Nagy:2023phz}.

To overcome these limitations, Non-Linear Integral Equation (NLIE) methods have been developed in various contexts 
\cite{KlumperPearce,Destri:1992qk,Klumper:1993vq,Destri:1994bv, Klumper:1992,en25}. 
These often allow for a more compact and efficient numerical treatment, as well as analytic control over the solutions. In this paper, 
we develop two such NLIE formulations, denoted as NLIE I and NLIE II, to describe the generalized Gibbs ensemble in the sine-Gordon model. 
Our construction starts from the TBA description of the GGE at special integer values of the coupling constant $p$, where the equations can be 
organized according to a ${\cal D}_{p+1}$ Dynkin-diagram structure \cite{Takahashi:1972zza,Zamolodchikov:1991et,Ravanini:1992fi}. 
We make use of the well-established techniques \cite{Jsuz} based on the $Y$-system and $T$-$Q$ relations  to reformulate the TBA equations into NLIEs \cite{BH1,BAR1,BAR2}. Although our derivation is performed at special points, the resulting equations admit analytic continuation to arbitrary values of the coupling constant in the repulsive regime. In fact, the dependence on the coupling is encoded entirely in the kernel functions of the integral equations, allowing for a straightforward generalization.
The two NLIE formulations have different domains of applicability. While NLIE I allows for limited extension into the attractive regime, it suffers from restrictions due to the shrinking analyticity strips of the unknown functions, which limits the number of conserved charges that can be included. In contrast, the NLIE II formulation is free from this limitation and permits an arbitrary number of charges to be coupled into the ensemble, but only in the whole repulsive regime.

Within their respective domains, our NLIE formulations provide an efficient and versatile framework for computing the generalized Gibbs potential, as well as expectation values of conserved charge densities, their associated currents, and local operators with respect to the GGE, at any value of 
the coupling constant of the theory. 
These results provide 
powerful tools for the analytical and numerical investigation of thermodynamic properties in the sine-Gordon model, both in and out of equilibrium. 


The plan of the paper is as follows. 
In section \ref{sectmodel}, we fix our conventions and summarize the TBA description of the GGE in the repulsive 
regime at the special points of the coupling constant, where the equations can be encoded into a ${\cal D}$-type Dynkin diagram. 
In section \ref{sectNLIE}, we  derive the two equivalent NLIE formulations for the TBA description of the GGE.
Section \ref{sectattr}, contains a discussion on the reason for the 
restricted validity of NLIE I in the attractive regime. 
 In section \ref{sectexp}, we  summarize,  how to compute expectation values of conserved charge densities, their associated currents and vertex operators in the context of our NLIE formalism. 
 Section \ref{sect6} closes the paper, with a short summary  and  some discussion of our results.

 The paper also contains an appendix. Appendix \ref{appA} provides numerical data supporting the validity of 
 the NLIE I formulation in the attractive regime of the model, in a suitable neighborhood of the free-fermion point.


\section{The model and the TBA equations} \label{sectmodel}

In this section, we summarize the thermodynamic Bethe ansatz (TBA) description of the generalized Gibbs ensemble \cite{Zamolodchikov:1989cf,Caux}  
in the sine-Gordon model at certain special points in the repulsive regime, where the equations become simply encoded into a ${\cal D}-$type 
Dynkin-diagram. 

The Euclidean action of the model is as follows: 
\begin{equation}
\label{sG_Lagrangian}
{S}_{_{SG}}[\phi]\!=\!\!\int \! dx\, d\tau  \left\{ \displaystyle\frac{1}{2} (\partial _{\mu }\phi(x,\tau))^2 -\displaystyle\frac{2\, \kappa^{2}}{\sin(\beta_{_{SG}} ^{2}/8)}\cos \left( \beta_{_{SG}} \, \phi(x,\tau) \right) \right\}\,  \qquad 0<\beta_{_{SG}}^2<8 \pi.
\end{equation}
In the sequel, we use the following parameterization of the coupling constant:
\begin{equation} \label{betap}
\begin{split}
p=\frac{\beta^2_{_{SG}}}{8 \, \pi -\beta^2_{_{SG}}}, \qquad \frac{\beta^2_{_{SG}}}{4\, \pi}=\frac{2 \, p}{p+1}, \qquad 0<p<\infty.
\end{split}
\end{equation}
The $p=1$  value  corresponds  to  the  free
fermion point of the theory, and the $0<p<1$ and $1<p$
regimes correspond to the attractive and repulsive regimes of the theory, respectively. 

We focus on computing the thermodynamic potential or free-energy density, 
derived from the partition function of the generalized Gibbs ensemble: 
\begin{equation} \label{fbetap}
\begin{split}
{\cal G}(\underline{\beta})=- \lim\limits_{L \to \infty}\frac{1}{L}\log {\cal Z}(\underline{ \beta}), 
\end{split}
\end{equation}
where 
\begin{equation} \label{Zbeta}
\begin{split}
{\cal Z}(\underline{ \beta})=\mbox{Tr}\left[ \exp\left({-\sum\limits_{j \in \mathbb{Z}} \beta_{2j-1} 
\hat{\cal H}_{2j-1}  } \right)\right],
\end{split}
\end{equation}
is the corresponding partition function, such that $\hat{\cal H}_{2j-1}$ are the dimensionless 
mutually commuting local integrals of motion with Lorentz spin
$2j-1.$ The parameter vector $\underline{\beta}=\{\beta_{2j-1} \}, \quad j \in \mathbb{Z},$ contains the dimensionless multipliers of these 
conserved quantities, which we call either potentials or generalized inverse temperatures. 
In the sine-Gordon model, the action of these dimensionless local conserved charges on the 1-particle soliton 
and anti-soliton states with rapidity $\theta:$ $|\theta,\pm \rangle$, is as follows: 
\begin{equation} \label{Hjact}
\begin{split}
\hat{\cal H}_{2j-1}\, |\theta,\pm \rangle= e^{(2 \, j-1) \theta} |\theta,\pm \rangle, \qquad j \in \mathbb{Z}.
\end{split}
\end{equation}
The usual Gibbs ensemble is a special case of (\ref{Zbeta}) with  $\beta_1=\beta_{-1}=\frac{\cal M}{2\,k_B \, T},$ 
such that ${\cal M}, \, k_B, T$ denote the soliton mass, the Boltzmann constant, and the temperature, respectively, and all the other 
potentials are zero ($\beta_j=0, \quad j\neq \pm 1$).

We consider the TBA equations coming from the saddle point condition for the partition function at the special points $2 \leq p \in \mathbb{N}.$ 
In this case the TBA equations can be encoded into a ${\cal D}_{p+1}$ Dynkin-diagram \cite{Zamolodchikov:1991et,Ravanini:1992fi,Tateo:1994pb}, in the following way:
\begin{equation} \label{TBAgge}
\begin{split}
\log y_j(x)&= -{\cal S}(x|\underline{\beta}) \, \delta_{j1}+\sum\limits_{k=1}^{N_p} \, I_{jk} \,(s*\log Y_k)(x), \qquad j=1,..,N_p=p+1, \\
Y_j(x)&=1+y_j(x).
\end{split}
\end{equation}
where $I_{jk}$ denotes the incidence matrix of the ${\cal D}_{N_p}$ Dynkin-diagram, $\delta$ stands for the Kronecker-delta symbol, 
\begin{equation} \label{TBAkernel}
\begin{split}
s(x)=\frac{1}{2 \, \pi \, \cosh (x)},
\end{split}
\end{equation}
is the standard TBA kernel, $*$ stands for convolution\footnote{In our convention:  $(f*g)(x)=\int\limits_{-\infty}^\infty dy \, f(x-y) \, g(y).$} and 
\begin{equation} \label{calS}
\begin{split}
{\cal S}(x|\underline{\beta})=\sum\limits_{k \in \mathbb{Z}} \beta_{2k-1} \, e^{(2 \, k-1)\, x},
\end{split}
\end{equation}
is the source term. Here, for later convenience we used the $Y-$system based notation for the unknowns of the TBA equations. 
In this formulation, the equations (\ref{TBAgge}) are well defined, if the series (\ref{calS}) converges  
for any $x \in \mathbb{R},$ and the function ${\cal S}(x|\underline{\beta})$ 
is bounded from below. Nevertheless, for practical computations, it is more realistic to consider the case, when 
only a finite number of higher-spin conserved charges are contained in the Gibbs ensemble. 
Thus, in our considerations, we focus on the case, when the number of higher-spin conserved charges is truncated. 
This corresponds only to a simple ${\cal S} \to {\cal S}_N$ change in the source term of the TBA equations (\ref{TBAgge}), where: 
\begin{equation} \label{calSN}
\begin{split}
{\cal S}_N(x|\underline{\beta})=\sum\limits_{k=1}^N   \bigg\{ \beta_{2k-1} \, e^{(2 \, k-1)\, x}
+\beta_{-(2k-1)} \, e^{-(2 \, k-1)\, x} \bigg\}.
\end{split}
\end{equation}
In this case, the requirement for ${\cal S}_N(x|\underline{\beta})$ is that $\lim\limits_{x \to \pm \infty }{\cal S}_N(x|\underline{\beta})\neq -\infty.$

Having the solution of the saddle point TBA equations (\ref{TBAgge}), two important thermodynamic potentials, the free-energy 
density ${\cal G}(\underline{\beta})$ (\ref{fbetap}), and the free-energy flux ${\Phi}(\underline{\beta})$ \cite{GHD1,Doyon}, 
can be computed by the formulas, as follows: 
\begin{equation} \label{EPtba}
\begin{split}
{\cal G}(\underline{\beta})=-{\cal M}  \intinf \!  \,  \frac{dx}{2 \pi} \, &\cosh\left(  x \right) \, \log Y_1 (x), \\
{\Phi}(\underline{\beta})=-{\cal M}  \intinf \!  \,  \frac{dx}{2 \pi} \, &\sinh\left(  x \right) \log Y_1 (x),
\end{split}
\end{equation}
where ${\cal M}$ denotes the soliton mass. 
They are generators of the expectation values of the dimensionless conserved charge densities and their associated currents, 
since their derivatives with respect to the inverse temperatures  give the  expectation values according 
to the formulas (\ref{qjpot}).  Due to the differentiating process, the explicit formulas for the 
expectation values contain the solutions of linear integral equations obtained by differentiating the 
generalized TBA equations (\ref{TBAgge}) with respect to the generalized inverse temperatures.

Our purpose is to reformulate the TBA equations (\ref{TBAgge})-(\ref{calSN}), which are valid only for 
some special values of the coupling constant,  
as compact few component non-linear integral equations that are valid for 
any value of the coupling constant in the repulsive regime. 
As we will see in the next section, one of the NLIE formulations (NLIE I) is found 
to be valid in a certain region of the attractive regime as well.

\section{The derivation of the NLIEs} \label{sectNLIE}

In this section, starting from the TBA equations presented in the previous section, 
we derive two different equivalent compact NLIE formulations of the problem.
Since, these equations depend analytically on the coupling constant, analytical continuation 
from the discrete values corresponding to the TBA formulation (\ref{TBAgge}) to the whole 
repulsive regime is straightforward.  

The derivation of both sets of equations 
are based on the method worked out earlier for the $O(3)$ non-linear $\sigma-$model 
\cite{BH1}, and for the sausage-model \cite{BAR1,BAR2}. The transformation of TBA to NLIE, relies on the integrable structure of the TBA equations, 
realized by the so-called $Y-$ and  $T-$system functional relations, and their solution through the famous $T-Q$ relations. 

As a first step, we collect all these necessary ingredients. 

\subsection{Functional relations stemming from the TBA}\label{func_sec}

The transformation of TBA to $Y-$system, starts with the observation  that the source functions (\ref{calS}) and (\ref{calSN}) are 
solutions to a kind of zero-mode functional equations: 
\begin{equation} \label{calSfunc}
\begin{split}
{\cal S}(x+i \, \tfrac{\pi}{2}|\underline{\beta})+{\cal S}(x-i \, \tfrac{\pi}{2}|\underline{\beta})=0, \qquad 
{\cal S}_N(x+i \, \tfrac{\pi}{2}|\underline{\beta})+{\cal S}_N(x-i \, \tfrac{\pi}{2}|\underline{\beta})=0.
\end{split}
\end{equation}
Then, it is straightforward to show \cite{Zamolodchikov:1991et} that the solution of TBA equations (\ref{TBAgge}) satisfy the $Y-$system functional equations:
\begin{equation} \label{Ysyst}
\begin{split}
y_1^+\, y_1^-&=Y_2, \\
y_k^+\, y_k^-&=Y_{k-1}\, Y_{k+1}, \qquad \qquad k=2,..,N_p-3,\\
y_{N_p-2}^+\, y_{N_p-2}^-&=Y_{N_p-3}\, Y_{N_p-1} \, Y_{N_p}, \\
y_{N_p-1}^+\, y_{N_p-1}^-&=y_{N_p}^+\, y_{N_p}^-=Y_{N_p-2},
\end{split}
\end{equation}
where we introduced the notation for any function $f^\pm(x)=f(x\pm i \, \pi /2).$ 
The first $N_p-3$ elements of these functions satisfy standard $su(2)$ type $Y-$system relations.  
To this part one can associate a $T-$system of the form\footnote{Here, the special, though natural $\phi=1$ gauge is used in the 
terminology of reference \cite{Wiegmann}.}:
\begin{equation} \label{Tsyst}
\begin{split}
T_k^+ \, T_k^-=1+T_{k-1} \, T_{k+1},  \qquad k=0,..,N_p-1, \qquad \mbox{with } \quad T_{-1}=0.
\end{split}
\end{equation}
The relations to the $Y-$functions is as follows:
\begin{equation} \label{YT}
\begin{split}
y_k&=T_{k-1}\, T_{k+1}, \qquad \qquad T_k^+ \, T_k^-=Y_k, \qquad k=1,..,N_p-2. \\
T_{N_p-1}^+ \, T_{N_p-1}^-&=Y_{N_p-1} \, Y_{N_p}.
\end{split}
\end{equation}
From the TBA equations (\ref{TBAgge}) and (\ref{YT}) it follows that both the $y-$ and $T-$ functions are real analytic. 
From (\ref{Ysyst}) and (\ref{YT}), it follows that:
\begin{equation} \label{ttT}
\begin{split}
y_{N_p-1}=y_{N_p}=T_{N_p-2}.
\end{split}
\end{equation}
From (\ref{Tsyst})-(\ref{ttT}), it can be shown \cite{BAR1} that:
\begin{equation} \label{TNTN2}
\begin{split}
T_{N_p}=2+T_{N_p-2}.
\end{split}
\end{equation}

For the $T-$system, one can associate the corresponding $T-Q$ system \cite{Wiegmann};
\begin{equation} \label{TQ}
\begin{split}
T_{k+1}  \, Q^{[k]}-T_k^-\, Q^{[k+2]}=\bar{Q}^{[-k-2]}, \qquad 
T_k^- \, \bar{Q}^{[-k]}-T_{k-1} \, \bar{Q}^{[-k-2]}=Q^{[k]}, 
\end{split}
\end{equation}
where we  introduced the notation $f^{[\pm k]}(x)=f(x\pm i \,k\,  \pi /2).$

Either $\bar{Q}$ or $Q$ can be eliminated from the $T-Q$ relations \cite{Wiegmann}, yielding second order 
difference equations:
\begin{equation} \label{diffQ}
\begin{split}
Q^{++}+Q^{--}=A\, Q \quad \mbox{with} \quad A=\frac{T_k^{[-k+1]}+T_{k-2}^{[-k-1]}}{T_{k-1}^{[-k]}},
\end{split}
\end{equation}
\begin{equation} \label{diffQbar}
\begin{split}
\bar{Q}^{++}+\bar{Q}^{--}=\bar{A}\, \bar{Q} \quad \mbox{with} \quad \bar{A}=\frac{T_k^{[k+3]}+T_{k+2}^{[k+1]}}{T_{k+1}^{[k+2]}},
\end{split}
\end{equation}
where both $A$ and $\bar{A}$ are $k-$independent. Thus, by choosing $k=N_p$ for $A$ and $k=N_p-2$ for $\bar{A},$ and using the identity 
(\ref{TNTN2}), one obtains:
\begin{equation} \label{AAbar}
\begin{split}
A=\frac{2+T_{N_p-2}^{[-N_p+1]}+T_{N_p-2}^{[-N_p-1]}}{T_{N_p-1}^{[-N_p]}}, \qquad \bar{A}=\frac{2+T_{N_p-2}^{[N_p-1]}+T_{N_p-2}^{[N_p+1]}}{T_{N_p-1}^{[N_p]}}.
\end{split}
\end{equation}
This implies together with (\ref{diffQ}) and (\ref{diffQbar}) the following relations \cite{BAR1}:
\begin{equation} \label{AAQQ}
\begin{split}
\bar{A}=A^{[2 \, N_p]}, \qquad \bar{Q}=Q^{[2 \, N_p]}.
\end{split}
\end{equation}
Though, the derivation of these relations is valid for positive integer values of $N_p,$ 
(\ref{AAQQ}) can be analytically continued for any real values of $N_p$ in  a straightforward manner.

\subsection{Analyticity strips} \label{anal_sec}

In order to transform the functional relations into integral equations, it is important to know about the functions,
their behaviour at infinity, and those analyticity  strips where they are free of zeroes and poles or any other type 
of singularities. This subsection is devoted to clarify these points for the functions introduced in the previous subsection. 

When we discuss, the analyticity strips, we consider the $y-, \, $ $T-, \,$ and $Q-$functions belonging to the TBA equations 
with source function (\ref{calSN}). Namely, we consider the case, when only a finite number of  higher conserved charges are 
coupled to the thermodynamic potential. We assume that the solution of  the TBA equations (\ref{TBAgge}), is a smooth deformation of the $\beta_j \to \infty$ limit. In this limit, the $y-,\,$ $T-,$ and $Q-$functions, tend to constant values, which we present below. 
The constant solution to the $y-$functions take the form:
\begin{equation} \label{yinf}
\begin{split}
y_k^{\infty}=(k-1)\,(k+1), \qquad k=1,...,N_p-2, \qquad y_{N_p}^{\infty}=y_{N_p-1}^{\infty}=N_p-2.
\end{split}
\end{equation}
The corresponding constant $T-$functions are:
\begin{equation} \label{Tinf}
\begin{split}
T_k^{\infty}=k, \qquad k=0,..., N_p-1.
\end{split}
\end{equation}
From this, also the functions $A$ and $\bar{A},$ can be computed in the infinite $\underline{\beta}$ limit:
\begin{equation} \label{Ainf}
\begin{split}
A^{\infty}=\bar{A}^{\infty}=2,
\end{split}
\end{equation}
which together with (\ref{diffQ}) and (\ref{diffQbar}) gives
\begin{equation} \label{Qinf}
\begin{split}
Q^{\infty}=\bar{Q}^{\infty}=1,
\end{split}
\end{equation}
provided one chooses the bounded solution of the difference equations \cite{BAR1}. 

From now on, we assume that we know the solution of the TBA equations (\ref{TBAgge}), when the 
$\beta_j$ constants are large, but finite. Then, all the important functions, namely, $y-, \quad T-,$ 
and $Q-$functions are smooth deformations of the infinite $\underline{\beta}$ solutions  
(\ref{yinf}), (\ref{Tinf}) and (\ref{Qinf}). Nevertheless, the infinite $\underline{\beta}$ solutions are not 
close to the solution of the TBA equations (\ref{TBAgge}) with large $\underline{\beta},$ on the whole complex plane. 
They are close to each other within finite strips in the imaginary directions. In the sequel, we will 
analyze the equations, and determine the strips, where the infinite $\underline{\beta}$ solution is close to the 
exact one with large, but finite  $\underline{\beta}.$ At the same time, these strips will be the analyticity strips 
mentioned at the beginning of this subsection. Namely, these are the strips, where the functions are analytic and free of any 
singularities and zeroes. 

Following the lines of \cite{BAR1,BAR2}, it is useful to introduce the $(\alpha_1,\alpha_2)$ short notation for the 
strip:
\begin{equation} \label{stripnot}
\begin{split}
\frac{\pi}{2}\, \alpha_1 < \text{Im}\, x < \frac{\pi}{2}\, \alpha_2.
\end{split}
\end{equation}
For later convenience, introduce the abbreviation ANZSC, which means that a function is analytic, non-zero, free of singularities, and 
tend to a constant value at infinity within a given strip of the complex plane. 

Now, we turn to the TBA equations (\ref{TBAgge}), and start to analyze the analyticity strips of the $y-$functions. 
First, the analyticity of $y_1$ should be considered. Due to the source term (\ref{calSN}): 
\begin{equation} \label{y1ANZ}
\begin{split}
y_1 \quad \text{is ANZSC in } \quad (-\epsilon_N,\epsilon_N) \quad \text{with} \quad \epsilon_N=\frac{1}{2N-1},
\end{split}
\end{equation}
which means that as the number of higher charges coupled to the thermodynamic potential increases, the analyticity strip for 
$y_1$ is shrinking. Using the standard techniques described in \cite{BH1} and \cite{BAR1,BAR2}, it can be shown that the TBA equations
(\ref{TBAgge}) and the $Y-$system relations imply the following analyticity strips for the $y-$functions:
\begin{equation} \label{yANZ}
\begin{split}
y_k \quad \text{is ANZSC in } \quad (1-k-\epsilon_N,k-1+\epsilon_N) \quad k=1,...,N_p-2.
\end{split}
\end{equation}
Following \cite{BH1,BAR1,BAR2}, the relations (\ref{Tsyst})-(\ref{TNTN2}) give  a one-unit wider strip for the $T-$functions: 
\begin{equation} \label{TANZ}
\begin{split}
T_k \quad \text{is ANZSC in } \quad (-k-\epsilon_N,k+\epsilon_N) \quad k=0,...,N_p-1.
\end{split}
\end{equation}
Using this, from the defining relation (\ref{AAbar}), one obtains the analyticity strip as follows for $A$ and $\bar{A}:$
\begin{equation} \label{AAstrip}
\begin{split}
A \quad &\text{is ANZSC in } \quad (3-\epsilon_N,2 \, N_p-3+\epsilon_N), \\
\bar{A} \quad &\text{is ANZSC in } \quad (-2 N_p +3-\epsilon_N,-3+\epsilon_N).
\end{split}
\end{equation}
Then, the difference equations (\ref{diffQ}) and (\ref{diffQbar}), give the analyticity strips for $Q$ and $\bar{Q}:$
\begin{equation} \label{QQstrip}
\begin{split}
Q \quad &\text{is ANZSC in } \quad (1-\epsilon_N,2 \, N_p-1+\epsilon_N), \\
\bar{Q} \quad &\text{is ANZSC in } \quad (-2 N_p +1-\epsilon_N,-1+\epsilon_N),
\end{split}
\end{equation}
which is consistent with the relation (\ref{AAQQ}) and the conjugation relation between $Q$ and $\bar{Q}.$

\subsection{Deriving NLIE I.}

The functional relations given in subsection \ref{func_sec} and the analyticity strips given in the previous subsection \ref{anal_sec} 
allow one to derive simple non-linear integral equations for computing the generalized Gibbs potential. 
To do so, the following functions composed out of the 
$y-,\,$ $T-,$ and $Q-$functions, turns out to be useful \cite{Jsuz,BH1,BAR1}. 
\begin{equation} \label{bBk}
\begin{split}
b_k&=\frac{Q^{[k+2]}\, T_k^-}{\bar{Q}^{[-k-2]}}, \qquad B_k=1+b_k=\frac{Q^{[k]}\, T_{k+1}}{\bar{Q}^{[-k-2]}}, \\
\bar{b}_k&=\frac{\bar{Q}^{[-k-2]}\, T_k^+}{Q^{[k+2]}}, \qquad 
\bar{B}_k=1+\bar{b}_k=\frac{\bar{Q}^{[-k]}\, T_{k+1}}{Q^{[k+2]}}, \qquad  k=0,..,N_p-2. 
\end{split}
\end{equation}
They satisfy the simple functional relations as follows:
\begin{equation} \label{bbbar}
\begin{split}
b_k \, \bar{b}_k&=Y_k, \qquad \qquad \qquad  B_k^+\, \bar{B}_k^-=Y_{k+1}, \qquad k=0,.., N_p-3, \\
b_{N_p-2} \, \bar{b}_{N_p-2}&=Y_{N_p-2}, \qquad  B_{N_p-1}^+\, \bar{B}_{N_p-1}^-=Y_{N_p-1}\, Y_{N_p}.
\end{split}
\end{equation}

In the rest of the paper, we  derive  two NLIEs. The first one is a Kl\"umper-Pearce-Batchelor-Destri-de Vega 
\cite{KlumperPearce,Destri:1992qk} type equation with a range of validity:  $1\!-\!\tfrac{1}{2\, N-1}<p.$ The second one 
is a kind of hybrid NLIE \cite{Jsuz,Dunning}, where the massive TBA function is kept as one of the unknown functions \cite{Heg1,Heg2}, 
and it is valid in the whole repulsive regime of the theory.


This subsection is devoted to the derivation of the first equation, which we call simply NLIE I.

The basic unknown functions of NLIE I, are the functions:
\begin{equation} \label{bb0}
\begin{split}
b_0&=\frac{Q^{[2]}\, T_0^-}{\bar{Q}^{[-2]}}, \qquad B_0=1+b_0=\frac{Q\, T_{1}}{\bar{Q}^{[-2]}}, \\
\bar{b}_0&=\frac{\bar{Q}^{[-2]}\, T_0^+}{Q^{[2]}}, \qquad \bar{B}_0=1+\bar{b}_0=\frac{\bar{Q}\, T_{1}}{Q^{[2]}},
\end{split}
\end{equation}
where $T_0$ is expressed in terms of the zero mode function:
\begin{equation} \label{T0}
\begin{split}
T_0(x)=e^{-{\cal S}_N(x|\underline{\beta})}, \qquad T_0^+\, T_0^-=1,
\end{split}
\end{equation}
with ${\cal S}_N(x|\underline{\beta})$ given in (\ref{calSN}). It is ANZSC in the strip $(-\epsilon_N,\epsilon_N).$ 

Now, it is important to discuss the ANZSC strips of the unknown functions (\ref{bb0}). 
The requirement of the existence of a finite width 
ANZSC strip, gives restrictions on the allowed range of the coupling constant $p$ defined in (\ref{betap}). 

First, we discuss the case of the repulsive regime, when there is no restriction. 
Consider $b_0$ in (\ref{bb0}). Using (\ref{TANZ})-(\ref{QQstrip}), one can list the analyticity strips 
of its building blocks:
\begin{equation} \label{b0block}
\begin{split}
Q^{[2]} \quad &\text{is ANZSC in } \quad (-1-\eps_N,2 \, N_p-3+\eps_N), \\
\bar{Q}^{[-2]} \quad &\text{is ANZSC in } \quad (-2 \, N_p+3-\eps_N,1+\eps_N), \\
T_0^- \quad &\text{is ANZSC } \quad (1 -\eps_N,1+\eps_N).
\end{split}
\end{equation}
In the repulsive regime, when $1<p$ and $2<N_p$ the intersection of these analyticity regions is 
$(1-\eps_N,1+\eps_N)$, which means that:
\begin{equation} \label{b0ANZ_rep}
\begin{split}
b_0^+ \quad \text{is ANZSC in } \quad (-\eps_N,\eps_N), \quad 1<p.
\end{split}
\end{equation}

In the attractive regime, when $0<p<1,$ there is nontrivial intersecting region of the strips (\ref{b0block}), 
if $1-\eps_N<2 \, N_p-3+\eps_N.$ This gives the restriction on the allowed range of $p,$ as follows:
\begin{equation} \label{prange}
\begin{split}
1-\eps_N<p<1, \qquad \eps_N=\frac{1}{2 \, N-1}. 
\end{split}
\end{equation}
In this case 
\begin{equation} \label{b0ANZ_attr}
\begin{split}
b_0^+ \quad \text{is ANZSC in } \quad (-\eps_N,\eps_N+2 \,(p-1)), \quad 1-\eps_N<p<1.
\end{split}
\end{equation}
Using (\ref{bb0}) a completely analogous train of thoughts leads to the following analyticity strips for 
the other unknowns: 
\begin{equation} \label{bbstrips}
\begin{split}
\bar{b}_0^- \quad &\text{is ANZSC in } \quad (-\eps_N,\eps_N), \qquad \qquad \qquad \qquad \quad 1<p<\infty, \\
\bar{b}_0^- \quad &\text{is ANZSC in } \quad (-\eps_N-2 \,(p-1),\eps_N), \qquad 1-\eps_N<p<1, \\
B_0^+, \, \bar{B}_0^- \quad &\text{is ANZSC in } \quad (-\eps_N,\eps_N), \qquad \qquad \qquad \quad 1-\eps_N<p<\infty.
\end{split}
\end{equation}
The ANZSC strips are important, because within these domains the logarithmic derivatives of the relevant 
functions{\footnote{The relevant functions are the $y-\,$ $T-,\,$ $Q-,$ and $b-$functions introduced above as part of the 
integrable hierarchy derived from the TBA equations.}}  can be Fourier-transformed, and the functional relations 
among them can be translated into algebraic equations between their Fourier-transforms \cite{KlumperPearce,Jsuz}. 

To carry out this procedure, first we fix our conventions for the Fourier-transformation. 
The Fourier-transformation rules between a function $f(x),$ and its transform  $\tilde{f}(\om),$ in our 
conventions are as follows:
\begin{equation} \label{FFtf}
\begin{split}
\tilde{f}(\om)=\intinf dx \, e^{i \, \om \, x} \, f(x), \qquad  
{f}(x)=\intinf \frac{dx}{2 \pi} \, e^{-i \, \om \, x} \, \tilde{f}(\om).
\end{split}
\end{equation}
Then, we introduce the notation $\dlk f$ for the Fourier-transform of the logarithmic-derivative of a function $f:$
\begin{equation} \label{dlf}
\begin{split}
\dlk f(\om)=\intinf dx \, e^{i \, \om \, x} \, \frac{d}{dx} \log f(x).
\end{split}
\end{equation}
The Fourier-transform of the TBA kernel (\ref{TBAkernel}) becomes:
\begin{equation} \label{sFFT}
\begin{split}
\tilde{s}(\om)=\frac{1}{2 \, \cosh \left(\tfrac{\pi \, \om}{2} \right)}.
\end{split}
\end{equation}
For later convenience, we denote:
\begin{equation} \label{qdef}
\begin{split}
q=e^{\tfrac{\pi \, \om}{2}}.
\end{split}
\end{equation}
If $\dlk f$ is ANZSC in the strip $(-K,K)$ with $0<K\in \mathbb{R},$  
then the following relation holds between the Fourier-transforms along 
different lines of the complex plane: 
\begin{equation} \label{fshiftrule}
\begin{split}
\dlk f^{[\pm k]}=q^{\pm k} \, \dlk f, \qquad    0<k<K.
\end{split}
\end{equation}

For practical purposes, it is convenient to introduce the  auxiliary function, as follows: 
\begin{equation} \label{a0}
\begin{split}
a_0=\frac{Q^{[2]}}{\bar{Q}^{[-2]}},
\end{split}
\end{equation}
in terms of which 
\begin{equation} \label{ba0}
\begin{split}
b_0=a_0\, T_0^-, \qquad  \bar{b}_0= \frac{T_0^+}{a_0}.
\end{split}
\end{equation}
From (\ref{b0block}) the analyticity strips for $a_0$ are as follows:
\begin{equation} \label{a0strips}
\begin{split}
a_0 \quad &\text{is ANZSC in } \qquad \qquad \qquad \quad (-1-\eps_N,1+\eps_N), \qquad 1<p<\infty.  \\
a_0 \quad &\text{is ANZSC in } \quad (-(2 \, p-1)-\eps_N,\eps_N+2 \,p-1), \quad 1-\eps_N<p<1.
\end{split}
\end{equation}
Now, we express $a_0$ in terms of $B_0$ and $\bar{B}_0$ within the appropriate analyticity strips. 
Since, the explicit form of $T_0$ is known, from such an equation the NLIE for $b_0$ and $\bar{b}_0$ can be 
easily derived with the help of the definitions (\ref{ba0}).

Now, we derive the algebraic equations in Fourier-space. 
Let,
\begin{equation} \label{q2}
\begin{split}
q_2=\dlk Q^{[2]},
\end{split}
\end{equation}
then from (\ref{AAQQ}) and the analyticity strips (\ref{QQstrip}):
\begin{equation} \label{dlQbarm2}
\begin{split}
\dlk \bar{Q}^{[-2]}=\dlk Q^{[2 \, N_p-2]}=q^{2 \, N_p-4}\, q_2.
\end{split}
\end{equation}
Using the definition (\ref{a0}) and the analyticity strips (\ref{a0strips}) for $a_0,$ one obtains:
\begin{equation} \label{a0FF}
\begin{split}
\dlk a_0=(1-q^{2\, N_p-4}) \, q_2.  
\end{split}
\end{equation}
From (\ref{bb0}), (\ref{q2}) and  (\ref{dlQbarm2}) the following two equations can be derived:
\begin{equation} \label{BB0FF}
\begin{split}
\dlk B_0^+=\frac{1}{q} \, q_2-q^{2 \, N_p-3} \, q_2+q\,  \dlk T_1, \\
\dlk \bar{B}_0^-=q^{2 \, N_p-3} \, q_2-\frac{1}{q} \, q_2 +\frac{1}{q}\,  \dlk T_1.
\end{split}
\end{equation}
These equations allow one to eliminate $T_1$ and $q_2$ and write them into (\ref{a0FF}). As a result, one obtains:
\begin{equation} \label{a0eq1}
\begin{split}
\dlk a_0=\frac{q^{N_p-2}-q^{2-N_p}}{(q^{N_p-1}-q^{1-N_p})\, (q+q^{-1})} \, ( q^{-1} \, \dlk B_0^+ -q\, \dlk \bar{B}_0^- ),  \qquad q=e^{\tfrac{\pi \, \om}{2}}. 
\end{split}
\end{equation}
Or equivalently in the Fourier-variable $\om:$
\begin{equation} \label{a0eq2}
\begin{split}
\dlk a_0(\om)=&G_p(\om) \, \left( e^{-\tfrac{\pi \, \om}{2}} \, \dlk B_0^+(\om) -e^{\tfrac{\pi \, \om}{2}}\, \dlk \bar{B}_0^-(\om) \right), \\
\quad  \text{with} \quad 
&G_p(\om)=\frac{\sinh\left((p-1)\, \tfrac{\pi \, \om}{2}\right)}{2 \, \cosh \left( \tfrac{\pi \, \om}{2} \right) \, \sinh\left(p\, \tfrac{\pi \, \om}{2}\right)}.
\end{split}
\end{equation}
Finally, one introduces the following new unknown functions:
\begin{equation} \label{newas}
\begin{split}
{\mathfrak a}_0(x)&=b_0^+(x-i \, \gamma \, \tfrac{\pi}{2}), \qquad \,  \, \bar{{\mathfrak a}}_0(x)=\bar{b}_0^-(x+i \, \gamma \, \tfrac{\pi}{2}), \\
{\mathfrak A}_0(x)&=B_0^+(x-i \, \gamma \, \tfrac{\pi}{2}), \qquad \bar{{\mathfrak A}}_0(x)=\bar{B}_0^-(x+i \, \gamma \, \tfrac{\pi}{2}), 
\end{split}
\end{equation}
where $\gamma$ is a contour deformation parameter, whose value is restricted by the analyticity domains of the variables (\ref{b0ANZ_rep})-(\ref{bbstrips}) to 
\begin{equation} \label{gamma}
\begin{split}
0< \gamma<\eps_N=\frac{1}{2 \, N-1}.
\end{split}
\end{equation}
We note that this range is valid in the repulsive regime, where the derivation of the equations is performed. 

The definitions (\ref{newas}) imply the trivial relations:
\begin{equation} \label{aU}
\begin{split}
{\mathfrak A}_0(x)=1+{\mathfrak a}_0(x), \qquad \bar{{\mathfrak A}}_0(x)=1+\bar{{\mathfrak a}}_0(x).
\end{split}
\end{equation}
Thus, the final form of the equations NLIE I, in rapidity space takes the form:
\begin{equation} \label{nlie1}
\begin{split}
\log {\mathfrak a}_0(x)&=-{\cal S}_N(x-i \tfrac{\pi}{2} \, \gamma)+(G_p * \log {\mathfrak A}_0)(x)-(G_p^{[2(1-\gamma)]} * \log \bar{{\mathfrak A}}_0)(x), \\
\log \bar{{\mathfrak a}}_0(x)&=-{\cal S}_N(x+i \tfrac{\pi}{2} \, \gamma)+(G_p * \log \bar{{\mathfrak A}}_0)(x)-(G_p^{[-2(1-\gamma)]} * \log {\mathfrak A}_0)(x). 
\end{split}
\end{equation}
These equations are derived at the special points of the coupling constant, when $2 \leq p \in \mathbb{N}.$ The final form of the NLIE I (\ref{nlie1}) 
is analytical in $p,$ providing a straightforward analytical continuation within the domain: $1-\tfrac{1}{2 \, N-1}<p<\infty.$ 
The lower bound comes from the fact that in the $p \leq 1-\tfrac{1}{2 \, N-1}$ regime, the width of the necessary analyticity strips would shrink to zero. 
This means that our NLIE I given in (\ref{nlie1}), offers only a shallow penetration into the attractive regime of the theory, such that the depth of 
the penetration is restricted by the highest spin entering the series of the  coupled higher-spin conserved charges.

In the attractive regime, the range of the contour deformation parameter $\gamma$ should also be modified with respect to that of the repulsive regime (\ref{gamma}), 
because in this regime the closest to the real axis poles of $G_p(x)$ are located at $\pm i \, p \, \pi,$ which should be avoided by the contour. This imposes the restriction: 
$(1-\gamma) \, \pi< p \, \pi,$ which implies the following range for $\gamma$ in the attractive regime:
$$1-p<\gamma<\eps_N=\tfrac{1}{2 N-1}.$$

Thus, the whole range of allowed values of $\gamma$ for  NLIE I (\ref{nlie1}), takes the form:
\begin{equation} \label{gamma_I}
\begin{split}
\text{max}(0,1-p)<\gamma<\eps_N, \qquad 1-\eps_N<p<\infty.
\end{split}
\end{equation}

 As a consequence of the relations (\ref{bbbar}), the TBA variables are related to those of the NLIE, as follows:
\begin{equation} \label{Y1AA}
\begin{split}
1+y_1=Y_1={\mathfrak A}_0^{[\gamma]} \, \bar{{\mathfrak A}}_0^{[-\gamma]}.
\end{split}
\end{equation}
This allows one to express the generalized Gibbs potential and the free-energy flux given in (\ref{EP}) in terms of our 
NLIE functions, as follows:
\begin{equation} \label{EP}
\begin{split}
{\cal G}(\underline{\beta})=-{\cal M}  \intinf \!   \frac{dx}{2 \pi} \,\bigg\{ &\cosh\left( x-i \, \tfrac{\pi}{2} \gamma \right) \, \log {\mathfrak A}_0 (x)+ 
\cosh\left( x+i \, \tfrac{\pi}{2} \gamma \right) \, \log \bar{{\mathfrak A}}_0 (x) \bigg\},  \\
\Phi (\underline{\beta})=-{\cal M}  \intinf \!  \frac{dx}{2 \pi} \,\bigg\{ &\sinh\left( x-i \, \tfrac{\pi}{2} \gamma \right) \, \log {\mathfrak A}_0 (x)+
\sinh\left(  x+i \, \tfrac{\pi}{2} \gamma \right) \, \log \bar{{\mathfrak A}}_0 (x)
\bigg\}.
\end{split}
\end{equation}
The formulas (\ref{nlie1}) together with (\ref{aU}), (\ref{gamma_I}) and (\ref{EP}) give the NLIE I formulation of the 
generalized thermodynamics.

The main disadvantage of this formulation, is that the allowed range for the contour deformation 
parameter becomes narrower and narrower as the value of $N$ increases. 
In this case, the difficulty comes from the fact that as the range of $\gamma$ tightens, the integration 
contour is forced to run closer and closer to the line, where infinitely many singularities of the unknown 
functions are condensated. In practice, if $N$ is large enough, numerical computations require taking very dense 
sampling point distributions.   


To avoid this scenario, in case the coupling constant is large enough, one can derive another set of hybrid NLIE equations, which keep the massive TBA 
node and sums up only the rest of the $y-$functions \cite{Heg1,Heg2}. This method works for $1<p<\infty,$ regime and we derive these equations in the next subsection. 


\subsection{Deriving NLIE II} 
 
In this section, we derive such an alternative hybrid-NLIE \cite{Jsuz,Dunning} 
description of the problem for $1<p,$ which keeps the contributions of the massive TBA node 
and sums up only the rest of the unknown functions of the TBA equations \cite{Heg1,Heg2}. 
 
Now, the complex unknown functions of the equations are the $k=1$ elements of the set of variables (\ref{bBk}):
\begin{equation} \label{bB1}
\begin{split}
b_1&=\frac{Q^{[3]}\, T_1^-}{\bar{Q}^{[-3]}}, \qquad B_1=1+b_1=\frac{Q^{[1]}\, T_{2}}{\bar{Q}^{[-3]}}, \\
\bar{b}_1&=\frac{\bar{Q}^{[-3]}\, T_1^+}{Q^{[3]}}, \qquad 
\bar{B}_1=1+\bar{b}_1=\frac{\bar{Q}^{[-1]}\, T_{2}}{Q^{[3]}}. 
\end{split}
\end{equation}
Using the results of the previous analyticity strip analysis (\ref{TANZ})-(\ref{QQstrip}), 
they have the following analyticity properties:
\begin{equation} \label{b1ANZ}
\begin{split}
b_1 &\text{ and } B_1: \quad \text{is ANZSC in } (-\eps_N,2+\eps_N), \\ 
\bar{b}_1 &\text{ and } \bar{B}_1: \quad \text{is ANZSC in } (-2-\eps_N,\eps_N).
\end{split}
\end{equation}
The unknown functions of the new set of equations called NLIE II, are $y_1,$ and the appropriately 
contour shifted versions of $b_1$ and $\bar{b}_1:$ 
\begin{equation} \label{newas1}
\begin{split}
{\mathfrak a}_1(x)&=b_1(x+i \, \gamma \, \tfrac{\pi}{2}), \qquad \,  \, \bar{{\mathfrak a}}_1(x)=\bar{b}_1(x-i \, \gamma \, \tfrac{\pi}{2}), \\
{\mathfrak A}_1(x)&=B_1(x+i \, \gamma \, \tfrac{\pi}{2}), \qquad \bar{{\mathfrak A}}_1(x)=\bar{B}_1(x-i \, \gamma \, \tfrac{\pi}{2}), 
\end{split}
\end{equation}
with $0<\gamma<1.$ Then, using the functional relations (\ref{Ysyst}), (\ref{YT}) and  (\ref{bbbar}), one can select the relevant ones:
\begin{equation} \label{rels2}
\begin{split}
y_1^+ \, y_1^-=Y_2, \quad B_1^+ \, \bar{B}_1^-=Y_2, \quad T_2^+ \, T_2^-=Y_2.
\end{split}
\end{equation}
Taking the logarithmic derivative of the relations (\ref{rels2}), one ends up  in Fourier-space with the following expression for $T_2$:
\begin{equation} \label{dlT2FF}
\begin{split}
\dlk T_2=\frac{1}{q+q^{-1}}\, \left( q \, \dlk B_1+ q^{-1} \, \dlk \bar{B}_1 \right).
\end{split}
\end{equation}
With the help of a straightforward analysis of the log of relations (\ref{rels2}), the following equation 
can be derived for $y_1:$ 
\begin{equation} \label{lny1}
\begin{split}
\log y_1(x)=\log T_0(x)+(s^{[1-\gamma]}*\log {\mathfrak A}_1)(x)+(s^{[\gamma-1]}*\log \bar{{\mathfrak A}}_1)(x),
\end{split}
\end{equation}
where we exploited (\ref{newas1}) and that $T_0$ is a zero mode function $T_0^+\, T_0^-=1.$ 

We just recall that around the real line, the exact functional form of $T_0$ is known: 
\begin{equation} \label{logT0S}
\begin{split}
\log T_0(x)=-{\cal S}_N(x|\underline{\beta}), 
\qquad -\eps_N<\text{Im}\, x <\eps_N, 
\end{split}
\end{equation}
with ${\cal S}_N$ given in (\ref{calSN}). 

To work in Fourier space, we denote:
\begin{equation} \label{q1def}
\begin{split}
q_1=\dlk Q^+.
\end{split}
\end{equation}
Then, due to (\ref{AAQQ}) and (\ref{QQstrip}):
\begin{equation} \label{q1bar}
\begin{split}
\dlk \bar{Q}^-=q^{2 \, N_p-2} \, q_1.
\end{split}
\end{equation}
Using the definitions (\ref{bB1}) together with the analyticity strip information (\ref{TANZ}) 
and (\ref{QQstrip}), one obtains the following equations among the Fourier-transforms of the logarithmic 
derivatives:
\begin{equation} \label{dlbB1}
\begin{split}
\dlk b_1&=q^2\, (1-q^{2\, N_p-6}) \, q_1+q^{-1}\, \dlk T_1, \\
\dlk B_1&= (1-q^{2\, N_p-4}) \, q_1+ \dlk T_2, \\
\dlk \bar{b}_1&=q^2\, (q^{2\, N_p-6}-1) \, q_1+q \, \dlk T_1, \\ 
\dlk \bar{B}_1&=q^2\, (q^{2\, N_p-4}-1) \, q_1+ \dlk T_2.
\end{split}
\end{equation}
From the relation $T_1^+ \, T_1^-=Y_1,$ coming from (\ref{YT}), one obtains in Fourier-space:
\begin{equation} \label{dlT1}
\begin{split}
\dlk T_1=\frac{1}{q+q^{-1}} \, \dlk Y_1.
\end{split}
\end{equation}
Eliminating $q_1,\, T_1$ and $T_2$ from the equations (\ref{dlT2FF}), (\ref{dlbB1}) and (\ref{dlT1}), one ends up with:

\begin{equation} \label{b1FFq}
\begin{split}
\dlk b_1=\frac{(q^{N_p-3}-q^{3-N_p})}{(q+q^{-1})(q^{N_p-2}-q^{2-N_p})} \, 
\left( \dlk B_1-\dlk \bar{B}_1 \right)+\frac{q^{-1}}{q+q^{-1}} \, \dlk Y_1,  \\
\dlk \bar{b}_1=\frac{(q^{N_p-3}-q^{3-N_p})}{(q+q^{-1})(q^{N_p-2}-q^{2-N_p})} \, 
\left( \dlk \bar{B}_1-\dlk B_1 \right)+\frac{q}{q+q^{-1}} \, \dlk Y_1.
\end{split}
\end{equation}
or equivalently in the Fourier variable $\om:$
\begin{equation} \label{b1om}
\begin{split}
\dlk b_1(\om)&=G_{p-1}(\om) \, \left(  \dlk B_1(\om)-\dlk \bar{B}_1(\om) \right)+
 \tilde{s}(\om) \, e^{-\tfrac{\pi \, \om}{2}} \, \dlk Y_1(\om), \\
\dlk \bar{b}_1(\om)&=G_{p-1}(\om) \, \left( \dlk \bar{B}_1(\om) -\dlk B_1(\om) \right)+
 \tilde{s}(\om) \, e^{\tfrac{\pi \, \om}{2}} \, \dlk Y_1(\om), 
\end{split}
\end{equation}
with $G_p(\omega)$ and $\tilde{s}(\omega)$ given in (\ref{a0eq2}) and (\ref{sFFT}), respectively. 

Finally, after integration and  turning to the variables defined in (\ref{newas1}), 
one ends up with the equations in rapidity space:
\begin{equation} \label{nlie2}
\begin{split}
\log {\mathfrak a}_1(x)&=(G_{p-1}*\log {\mathfrak A}_1)(x)-(G_{p-1}^{[2 \, \gamma]}*\log \bar{{\mathfrak A}}_1)(x)+
(s^{[\gamma-1]}*\log Y_1) (x), \\
\log \bar{{\mathfrak a}}_1(x)&=(G_{p-1}*\log \bar{{\mathfrak A}}_1)(x)
-(G_{p-1}^{[-2 \gamma]}*\log {\mathfrak A}_1)(x)+
(s^{[1-\gamma]}*\log Y_1) (x), \\
\log y_1(x)&=\log T_0(x)+(s^{[1-\gamma]}*\log {\mathfrak A}_1)(x)+(s^{[\gamma-1]}*\log \bar{{\mathfrak A}}_1)(x),
\end{split}
\end{equation}
where $0<\gamma<1.$ Though, the equations were derived only at the positive integer values of the coupling constant $p=2,3,...,$ they can be continued analytically to the whole range, $2<p$ in a straightforward manner.

It turns out that these equations can be further continued in $p$ even to the  $1<p<2$ regime. The derivation of the equations 
was based on the analyticity strips justified for $2<p.$ Now, having the equations at hand, they can be continued to the $1<p<2,$ regime, 
as well. Only, the kernel $G_{p-1}$ contains the $p-$dependence, thus the analytical continuation to $1<p<2$ seems to be straightforward. 
The only subtlety comes from the allowed range of $\gamma.$ Namely, the closest to the real axis poles of $G_{p-1}(x)$ at $\pm i \,(p-1)\, \pi$ 
should be avoided by the contour, which implies, the $0<\gamma<p-1$ range.  In summary, the allowed range for the contour deformation 
parameter $\gamma$ in NLIE II (\ref{nlie2}) is as follows:
\begin{equation} \label{gamma_II}
\begin{split}
0<\gamma<\text{min}(1,p-1), \qquad \quad 1<p<\infty.
\end{split}
\end{equation}

The formulas for the generalized Gibbs potential and the free-energy flux remain the same as in the TBA 
formulation (\ref{EPtba}):
\begin{equation} \label{EP2}
\begin{split}
{\cal G}(\underline{\beta})=-{\cal M} \intinf \! \,  \frac{dx}{2 \pi} \, &\cosh\left(  x \right) \, \log Y_1 (x), \\
{\Phi}(\underline{\beta})=-{\cal M}  \intinf \!  \,  \frac{dx}{2 \pi} \, &\sinh\left(  x \right) \log Y_1 (x).
\end{split}
\end{equation}

We close this section, with a short discussion on the applicability and benefits of the two NLIE formulations. 

NLIE I, has the main advantage that it is valid in the whole repulsive regime of the theory, and depending on 
the value of the spin of the largest spin conserved charge contained in the Gibbs potential, it can be valid 
in a certain neighborhood of the free-fermion point in the attractive regime, as well. 
Nevertheless, the main disadvantage of this formulation is that the analyticity strips for the unknown functions 
become very narrow, when conserved charges with high spin are present in the Gibbs ensemble. This implies that 
one has to integrate very closely to the infinitely many singularities of the unknown functions, which require 
very dense distribution of the sampling points, when numerical computations are done. 
Another issue arising from the same source is that when all higher-spin conserved quantities are coupled to 
the thermodynamic potential, the analyticity strip of the unknowns shrinks to zero. 
This implies that the equations are no longer applicable in this regime. 

As for NLIE II, in this case, there is no problem with neither the analyticity strips nor the number of contained 
higher spin conserved quantities. The only restriction of applicability is the range of coupling constant. 
It is narrower, than that of the NLIE I, and looses efficiency as the free-fermion point $p=1$ is closely approached. 

Nevertheless, when only a few extra higher spin conserved quantities are coupled to the thermodynamic potential, 
NLIE I, offers a still satisfactory description for the thermodynamics, but in a wider range of coupling constant.

In the next section, 
we discuss, from the point of view of the particle content of the theory, 
the reason why these restrictions on the validity of NLIE I arise in the attractive regime.

\section{Discussion of the attractive regime} \label{sectattr}

This section explores how the particle spectrum of the theory accounts for the restricted validity 
of NLIE I in the attractive regime.


Let us denote, the  dimensionless spin-$s$ conserved charges by $\hat{\cal H}_{s}$. 
Their action on one-particle soliton (+) or anti-soliton (-) states with rapidity $\theta$ are, as follows:
\begin{equation} \label{calHse}
\begin{split}
\hat{\cal H}_{\pm s} | \theta, \pm \rangle= {\cal H}_{\pm s} (\theta) \, | \theta, \pm \rangle 
=e^{\pm s \, \theta} \, | \theta, \pm \rangle, \qquad s=1,3,5,....,
\end{split}
\end{equation}
where the spin can take any odd integer value in the sine-Gordon model. 

The general formula for the partition function (\ref{Zbeta}) can be interpreted as the partition function of a 
usual Gibbs ensemble, but with a generalized Hamiltonian depending on the generalized inverse temperatures:
\begin{equation} \label{Hgen}
\begin{split}
Z(\underline{\beta})&=\text{Tr} \, e^{-\beta \, \hat{\cal H}_{gen}(\underline{\beta})}, \\
\hat{\cal H}_{gen}(\underline{\beta})&=\frac{1}{\beta}\sum\limits_{j \in \mathbb{Z}} \beta_{2j-1} \,
\hat{\cal H}_{2j-1}, \qquad \beta=\frac{1}{k_B \, T}.
\end{split}
\end{equation}
In order for the partition function to exist, $\hat{\cal H}_{gen}$ must be a lower bounded operator. 
In the repulsive regime, when only solitons and anti-solitons form the spectrum of the theory, 
the eigenvalues of higher-spin conserved charges on one-particle states are positive. Namely, they are bounded from below. 
Thus, if all inverse temperatures are positive, the generalized Hamiltonian becomes automatically bounded from below.


If one starts to penetrate into the attractive regime by decreasing the value of $p$, the first breather enters the spectrum. 
If we restrict our attention to the range $1/2<p<1,$ only the first breather, the soliton and the anti-soliton form the 
spectrum of the theory. Consequently, in the partition function (\ref{Zbeta}), 
the breather contributions will arise, as well. The 
corresponding one-particle eigenvalues, can be computed using the bootstrap principle \cite{zamzam79}. 
\begin{equation} \label{calHB1}
\begin{split}
\hat{\cal H}_{\pm s}\,| \theta, B_1 \rangle&={\cal H}_{B_1,\pm s}(\theta) \,| \theta, B_1 \rangle, \\
{\cal H}_{B_1,\pm s}(\theta)&={\cal H}_{\pm s}\left(\theta+i \, \tfrac{\pi}{2} \left(1-\tfrac{p}{2}\right)\right)+
{\cal H}_{\pm s}\left(\theta-i \, \tfrac{\pi}{2} \left(1-\tfrac{p}{2}\right)\right), \\
{\cal H}_{B_1,\pm(2n-1)}(\theta)&=(-1)^{n+1} \, 2\, \sin \left[ (2\, n-1)\,\frac{p \, \pi}{2} \right] \, e^{ \pm (2 \,n-1) \, \theta},
\qquad n=1,2,...
\end{split}
\end{equation}
where ${\cal H}_{B_1,\pm s}(\theta)$ and ${\cal H}_{\pm s}(\theta)$ are the $B_1$-breather and 
the soliton or anti-soliton eigenvalues of 
the higher spin conserved quantities on a one-particle state with rapidity $\theta$. As we mentioned earlier, for the soliton and anti-soliton 
states these eigenvalues are always bounded from below in $\theta.$ Unfortunately, this doesn't hold for any value of the coupling constant 
for the $B_1$ case. It turns out that ${\cal H}_{B_1,\pm s}(\theta)$ is bounded from below, only if:
\begin{equation} \label{boundB1}
\begin{split}
1-\frac{1}{s}<p, \qquad \qquad s>0.
\end{split}
\end{equation}
As the spin increases, the range in which the conserved charges remain bounded from below becomes smaller and smaller. 
If only the first \( N \) higher-spin conserved charges are included in the generalized 
Hamiltonian{\footnote{This scenario corresponds to the source term \( \mathcal{S}_N \) (\ref{calSN}) in the NLIE or TBA formalism.}} (\ref{Hgen}), then the condition ensuring that all contributing conserved charges are bounded from below is:
\begin{equation} \label{boundN}
\begin{split}
1 - \frac{1}{2N - 1} < p.
\end{split}
\end{equation}
This is also the range of validity for NLIE I (\ref{nlie1}). Within this range, the generalized Hamiltonian (\ref{Hgen}) is guaranteed to be bounded from below provided the generalized inverse temperatures are positive.  However, it may still be bounded from below even when the condition (\ref{boundN}) is not satisfied. The key point is that NLIE I requires a stronger condition: not just the full Hamiltonian, but each individual higher-spin contribution must be bounded from below.

This distinction matters, as it limits the applicability of NLIE I beyond the general requirement that 
$\hat{\cal H}_{gen} $ be bounded from below. To illustrate this, consider a simple example where the generalized Hamiltonian is:
\begin{equation} \label{Hexample}
\begin{split}
\hat{\cal H}_{\text{gen}} = \hat{\cal H}_3 + \beta^{\prime}_9 \, \hat{\cal H}_9, \qquad \text{with} \quad 0 < \beta^{\prime}_9.
\end{split}
\end{equation}
It is easy to show from (\ref{calHB1}) that this Hamiltonian is bounded from below for one-particle \( B_1 \) states in the following range of the coupling constant:
\begin{equation} \label{bfb'}
\begin{split}
\frac{4}{9} < p < \frac{2}{3} < 1 - \frac{1}{9}.
\end{split}
\end{equation}
Interestingly, this lies outside the NLIE I validity region, which requires \( 1 - \frac{1}{9} < p \). This example clearly shows that NLIE I cannot account for all cases where the generalized Hamiltonian is bounded from below. It only works in the more restrictive domain where each contributing higher-spin conserved charge is individually bounded from below.

However, we are not claiming that no alternative to NLIE I exists in the regions where it fails. It is possible that a more refined approach than the Thermodynamic Bethe Ansatz (TBA) can be developed. What we emphasize is that the straightforward extension of the NLIE method, as presented in \cite{KlumperPearce,Destri:1992qk,Klumper:1993vq,Destri:1994bv}, is no longer valid in those regions. To describe the thermodynamics of the model accurately, a more sophisticated set of equations would be necessary.


\section{Expectation values} \label{sectexp}

In this section, we discuss, how the NLIE I and II formulations allow one to compute expectation values of important operators in the theory. 

\subsection{Expectation values of conserved charges and their associated currents}

The averages of the local conserved charge densities{\footnote{The connection with the notation introduced in (\ref{Zbeta}) is $\hat{\cal H}_i=\int \! dx \, {\bf q}_i(x,t)$.}}
 ${\bf q}_i$ and their associated currents ${\bf j}_i$ can be straightforwardly computed from both NLIE I and II, as the generating potentials of these quantities can be directly expressed in terms of the solutions of the NLIEs.


In the sequel, we give formulas for the averages of dimensionless charge densities and  their  
associated currents. They can be computed by differentiating the 
corresponding free-energy densities \cite{GHD1,Doyon}:
\begin{equation} \label{qjpot}
\begin{split}
{\bf q}_i=\langle q_i(x,0)\rangle_{\beta}=\frac{\partial}{\partial \beta_i} {\cal G}(\underline{\beta}), \\
{\bf j}_i=\langle j_i(x,0)\rangle_{\beta}=\frac{\partial}{\partial \beta_i} \Phi(\underline{\beta}),
\end{split}
\end{equation}
where the potentials ${\cal G}$ and $\Phi$ are given in (\ref{EP}) and (\ref{EP2}) for the NLIE I and II formulations, respectively.
As usual, after the differentiation, the averages are expressed in terms of solutions to linear integral equations obtained by differentiating the NLIE I (\ref{nlie1}) or II (\ref{nlie2}) equations.

\subsection{Expectation values of local operators}

Beyond computing the expectation values of conserved charges and  the associated currents, our NLIE formalism allows one to compute  the expectation values of local operators, as well. 

Particularly, the expectation values of the vertex operators $e^{\pm i \, m \, \beta_{{}_{SG}} \,\phi(x)}$ in the sine-Gordon model, with $m \in \mathbb{N},$ 
become important,  since they correspond to  experimentally measurable coherence factors in tunnel-coupled cold atomic condensates 
\cite{Hofferberth:2007qfs,Esslernincs}. To keep the formulas as short as possible, we introduce the notation:
$$ \nu=\frac{1}{p+1}.$$

The discovery of fermionic basis  \cite{Boos:2006mq,Boos:2008rh,Jimbo:2008kn,Boos:2010qii} in the 6-vertex model, and its application to 
the sine-Gordon model \cite{Jimbo:2010jv} makes the calculation of the vertex operator expectation values possible.

The expectation values of vertex operators  $e^{i \, m \, \beta_{_{SG}} \,\phi(x)},$ take the form, as follows  
 \cite{Jimbo:2010jv,Hegedus:2019rju}: 
\begin{equation}  \label{primVEV}
\begin{split}
\langle  e^{i \, m \, \beta_{{}_{SG}} \,\phi(x)} \rangle=
{\kappa}^{-2 \, m^2\,(1-\tfrac{1}{\nu})}\, C_m(0) \, \underset{1 \leq j,k \leq m }{\mbox{det}} \, \Omega_{kj}, \qquad m=1,2,...
\end{split}
\end{equation}
where for $m>0,$ $C_m(\alpha)$ is  given by:
\begin{equation} \label{Cm}
\begin{split}
C_m(\alpha)=\prod\limits_{j=0}^{m-1} C_1(\alpha+2j\tfrac{1-\nu}{\nu}),
\end{split}
\end{equation}
with
\begin{equation} \label{C1}
\begin{split}
&C_1(\alpha)=i \, \nu \Gamma(\nu)^{4 x(\alpha)}\, \frac{\Gamma(-2 \nu x(\alpha))}{\Gamma(2 \nu x(\alpha))}\, \frac{\Gamma(x(\alpha))}{\Gamma(x(\alpha)+1/2)} \,
\frac{\Gamma(-x(\alpha)+1/2)}{\Gamma(-x(\alpha))} \cot(\pi x(\alpha)), \\
&\text{and} \qquad x(\alpha)=\frac{\alpha}{2}+\frac{1-\nu}{2 \nu}.
\end{split}
\end{equation}
The matrix elements of $\Omega_{j,k}$ take the form,  as follows: 
\begin{equation}  \label{OM}
\begin{split}
\Omega_{j,k}=\om_{2k-1,1-2j}+\frac{i}{\nu} \, \delta_{j,k}\, 
\cot \left[ \frac{\pi}{2 \nu} (2k-1)\right], \qquad j,k=1,...m,
\end{split}
\end{equation}
where the matrix elements $\om_{2k-1,1-2j}$ involve the $\underline{\beta}$ dependence of the problem and are the fundamental building blocks of the 
formula for expectation values of vertex operators.

The parameter $\kappa$ in (\ref{primVEV}), is the coupling constant in the sine-Gordon action (\ref{sG_Lagrangian}). 
It is related to the soliton mass ${\cal M},$ by the  mass gap formula  \cite{Zamolodchikov:1995xk}:
\begin{equation} \label{muM}
\begin{split}
\kappa={\cal M}^\nu \, \Pi(\nu)^\nu, \qquad \text{where} \qquad 
\Pi(\nu)=\frac{\sqrt{\pi}}{2} \frac{\Gamma\left(\tfrac{1}{2 \nu}\right)}{\Gamma\left(\tfrac{1-\nu}{2 \nu}\right)} \, \Gamma\left( \nu \right)^{-\tfrac{1}{\nu}}.
\end{split}
\end{equation}

As for the formulas for  $\om_{2k-1,1-2j},$ we present them both in the NLIE I and NLIE II formulations.
To keep the formulas as short as possible, we present these matrix elements, as derivatives of appropriate generating functions with respect to 
the elements of the $\underline{\beta}$ vector. 
\begin{equation}  \label{omtom}
\begin{split}
\om_{2k-1,1-2j}=\frac{(-1)^k}{\pi \, \nu\, i}\,  \frac{\partial}{\partial \beta_{1-2\, j}} J_{2k-1}(\underline{\beta}),
\end{split}
\end{equation}
with 
\begin{equation} \label{J2km1}
\begin{split}
J_{2k-1}(\underline{\beta})&=\! \! \intinf dx \, \left(e_{2k-1}^{[-\gamma]}(x) \, \log {\mathfrak A}_0(x)+  e_{2k-1}^{[\gamma]}(x) \, \log \bar{{\mathfrak A}}_0(x)
\right), 
\quad \text{with NLIE I,} \\
J_{2k-1}(\underline{\beta})&=\! \! \intinf dx \, e_{2k-1}(x) \, \log Y_1(x), \qquad \qquad \qquad \quad \quad 1<p,\quad \text{with NLIE II,} 
\end{split}
\end{equation}
where $e_{2k-1}(x)=e^{(2k-1)x},$ and in case of NLIE I, the allowed range of $\gamma$ and $p$ is given in (\ref{gamma_I}).

Taking the explicit differentiations in (\ref{J2km1}), lead to expressions containing the solutions of specific  linear integral equations, obtained by differentiating the NLIE I (\ref{nlie1}) and II  (\ref{nlie2}) equations.



\section{Summary and discussion} \label{sect6}

In this paper we proposed two sets of nonlinear integral equations (NLIE) for 
describing the thermodynamics in the sine-Gordon model, when higher spin conserved charges are also coupled to 
the Gibbs ensemble. The first one, called NLIE I formulation is a direct extension of the 
Kl\"umper-Pearce-Bathchelor-Destri-de Vega type \cite{KlumperPearce,Destri:1992qk} equations to the case, when higher spin 
conserved charges are also included in the partition function of the Gibbs ensemble.  

The second one, called NLIE II, is a kind of hybrid-NLIE \cite{Jsuz,Dunning} being valid only in the repulsive regime, where 
the massive $Y-$function of the TBA formulation is kept and only the rest of the complicated TBA 
system is summed up by a few complex functions \cite{Heg1,Heg2}.  

The derivations of both sets of equations, are based on $T-Q$ relations given by the equivalent thermodynamic 
Bethe ansatz (TBA) formulation of the problem in the repulsive regime. Though the equations are derived in the 
repulsive regime, by appropriate analytical continuation, in case of  NLIE I, the penetration into the attractive 
regime is also possible. However, the magnitude of this penetration is restricted by the spin of the 
largest spin conserved charge contained in the Gibbs ensemble (\ref{gamma_I}). 

The main advantage of both sets of equations is that they are simple few component equations and their 
coupling dependence is encoded into the kernel, which allows simple analytical continuation in the 
coupling constant within their range of validity.  

Both sets of equations offer efficient frameworks for computing expectation values of local conserved 
densities, their associated currents, and vertex operators and their descendants, with respect to the 
generalized Gibbs ensemble.


Though, both sets of equations give the correct description of the generalized thermodynamics in the 
sine-Gordon model, both formulations have their own advantages and limitations. Thus, for specific purposes 
one should choose the appropriate formulation. 

The main restriction for the equations come from the shrinking of analyticity domains, when the coupling 
constant $p$ is approaching or going below the free-fermion point ($p=1)$, or when more and more higher spin 
conserved charges are coupled to the Gibbs ensemble.  

As a summary, it can be stated that NLIE II is a perfect formulation of the problem in the whole repulsive 
regime $p>1,$ any number of higher conserved charges can be coupled to the Gibbs ensemble. Expectation values 
of local densities, their associated currents and vertex operators $e^{\pm i\,m\, \beta_{_{SG}} \phi}$ and their descendants 
can be computed without any difficulty. Some care is needed, only when the coupling constant gets very close 
to the free-fermion point $p=1,$ since in this case the integration contour runs close to the first pole of the 
kernel $G_{p-1}(x),$ which requires more sampling points, when numerical integrations are carried out.  

As for NLIE I, there is  the shrinking of analyticity domains, as the spin of the largest spin conserved charge 
contained in the generalized Gibbs ensemble, increases.  Because of this phenomenon, within this formulation 
only a finite number of higher spin local conserved charges can be coupled to the Gibbs ensemble. 
If the largest spin is $2 \, N-1,$ then the range of validity in the coupling constant is given by 
(\ref{gamma_I}), 
$$1- \frac{1}{2\, N-1}<p<\infty.$$
Thus, this description can be still valid (beyond the repulsive regime)  
in the attractive regime, in certain neighborhood of the 
free-fermion point. Within its range of validity, it offers an efficient framework for 
computing the expectation values 
of local densities, their associated currents and vertex operators $e^{\pm i\,m\, \beta_{_{SG}} \, \phi}$ and their descendants.  

Though, not discussed in this paper, NLIE I formulation perfectly fits into the framework of \cite{Jimbo:2010jv} 
 for 
computing ratios of expectation values of local operators. Thus, NLIE I contrary to NLIE II, can describe 
ratios of expectation values of the extended set of vertex operators corresponding to the perturbed 
complex Liouville CFT description of the sine-Gordon model \cite{Jimbo:2010jv}. 
Namely, for example the ratios of expectation values 
$$\frac{\langle e^{\pm i \, (m\,+\alpha) \, \beta_{_{SG}}\, \phi} \rangle}{\langle e^{\pm i \alpha\, \beta_{_{SG}} \phi} \rangle}, 
\qquad \qquad m \in \mathbb{Z}, \qquad \qquad 0<\alpha<\frac{1}{p},$$ 
can also be computed by simply substituting into the formulas of references \cite{Jimbo:2010jv,Hegedus:2019rju}, but now changing 
the NLIE function to that of NLIE I, and applying a large enough value for the contour deformation parameter.

\vspace{1cm}
{\tt Acknowledgments}

\noindent 
The authors thank J\'anos Balog and Zolt\'an Bajnok for useful discussions.
This research was supported  by the NKFIH grant K134946.

\appendix

\section{Some numerical results} \label{appA}

This appendix is devoted to collecting numerical data for demonstrating that the NLIE I 
description  remains valid in the attractive regime, provided the coupling constant 
remains within a suitable neighborhood of the free fermion point (\ref{gamma_I}).

\begin{figure}
\begin{flushleft}
\hskip 25mm
\leavevmode
\epsfxsize=100mm
\epsfbox{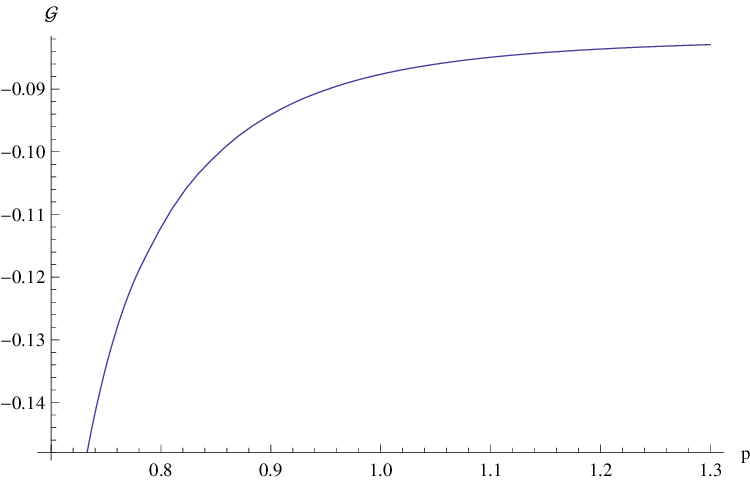}
\end{flushleft}
\caption{\footnotesize
The coupling constant dependence of the Gibbs potential around the free fermion point, when 
$\beta_1=3 \cdot 10^{-1},  {\beta}_{-1}= 10^{-1},   \beta_3=5/6,\, {\beta}_{-3}=1/6$.
}
\label{pdepGibbs}
\end{figure}

Table \ref{text1}, contains some numerical data for the coupling constant dependence of the Gibbs potential around the 
free fermion point, when at maximum spin-3 conserved charges are coupled to the thermodynamic potential and the soliton mass ${\cal M}$ 
is set to be $1$. 
The plot \ref{pdepGibbs}, gives the visual presentation of the same set of numerical data. 
In view of these data, the continuity in the coupling constant is plausible. 

Nevertheless, to give more evidence on the continuous transition between the repulsive and attractive regimes, we 
numerically computed the first two derivatives of the thermodynamic potential at the free fermion point in two ways.  
First from numerical data corresponding to the repulsive regime, and then from data coming from the attractive regime, and 
we checked that the numerical values of the first two derivatives do not depend on the regime where we computed them. This is a very strong numerical evidence of the correctness of the NLIE I description in the attractive regime.

Table \ref{text2}, contains the numerical data for the computation of the numerical derivatives. 
We used higher order formulas for the numerical computations of the derivatives.

Let $h$ be a small positive number used for computing the derivatives numerically, and let $f(\underline{\beta}|p)$ 
be an arbitrary function of the vector of generalized inverse temperatures and the coupling constant $p.$  
Denote:  
$$f_j=f(\underline{\beta}|1+j \, h), \qquad j=\pm1, \pm2, \pm 3,....$$  
These are the  discrete values of the function from which the numerical derivatives around the $p=1$ free-fermion point are computed.



We note that in our actual numerical computations we choose $h=10^{-3}$  for the value of $h.$ 
Thus, the data in table \ref{text2}, was generated by this choice for $h.$

Then the first numerical derivative at $p=1,$ can be computed by the following 2nd order formulas. 
Repulsive regime:
$$ \partial_p f(\underline{\beta}|p)|_{p=1}=-\frac{3\, f_0-4 \, f_1+f_2}{2 h}+O(h^2),$$
Attractive regime:
$$ \partial_p f(\underline{\beta}|p)|_{p=1}=-\frac{-3\, f_0+4 \, f_{-1}-f_{-2}}{2 h}+O(h^2).$$
Using the data given in table \ref{text2}, one obtains for the derivatives of ${\cal G}$:
$$ \partial_p {\cal G}(\underline{\beta}|p)|_{p=1}^{repulsive}=0.039615759858448, $$ 
$$ \partial_p {\cal G}(\underline{\beta}|p)|_{p=1}^{attractive}=0.039615739913836,$$ 
with the difference between the two computations 
$$(\partial_p {\cal G}(\underline{\beta}|p)|_{p=1}^{repulsive}-\partial_p {\cal G}(\underline{\beta}|p)|_{p=1}^{attractive})/\partial_p {\cal G}(\underline{\beta}|p)|_{p=1}^{repulsive}\sim 10^{-7}.$$ 
The second derivatives can be obtained similarly. 

The second numerical derivative at $p=1,$ can be computed by the following 2nd order formulas. 
\newline
Repulsive regime:
$$ \partial_p^2 f(\underline{\beta}|p)|_{p=1}=-\frac{-2\, f_0+5 \, f_1-4 \, f_2+f_3}{2 h^2}+O(h^2),$$
Attractive regime:
$$ \partial_p^2 f(\underline{\beta}|p)|_{p=1}=-\frac{-2\, f_0+5 \, f_{-1}-4\, f_{-2}+f_{-3}}{2 h^2}+O(h^2).$$
Using the data given in table \ref{text2}, one obtains:
$$ \partial^2_p {\cal G}(\underline{\beta}|p)|_{p=1}^{repulsive}=-0.16828915351411, $$ 
$$ \partial^2_p {\cal G}(\underline{\beta}|p)|_{p=1}^{attractive}=-0.168288637850,$$ 
with the difference between the two computations 
$$(\partial_p^2 {\cal G}(\underline{\beta}|p)|_{p=1}^{repulsive}-\partial_p^2 {\cal G}(\underline{\beta}|p)|_{p=1}^{attractive})/\partial_p^2 {\cal G}(\underline{\beta}|p)|_{p=1}^{repulsive}\sim 10^{-6}.$$ 
These results are consistent with the $O(h^2) \sim 10^{-6}$  error dictated by the second order numerical derivative formulas.

These numerical computations provide strong evidence for the validity of NLIE I in the attractive regime 
sufficiently close to the free fermion point (\ref{gamma_I}).  Using numerical derivative computations 
on both sides of the free-fermion point, 
we were able to show with high precision that, at this point,  
not only the curve (obtained from the solutions of NLIE I) in figure \ref{pdepGibbs}, but also 
at least its first and second derivatives are continuous.   


\begin{table}[h]
\begin{center}
\begin{tabular}{|c|c|c|c|}
\hline
$ p $ & $ {\cal G}(\underline{\beta}|p) $  & $\mbox{num. error}$   \tabularnewline
 \hline
 $0.7$  & $-0.187239866140983$  &   $10^{-6}$\\
 \hline
 $0.75$  & $-0.13423191891381$  &   $10^{-6} $\\
 \hline
 $0.79$  & $-0.11544334423915$  &   $10^{-8}$\\
 \hline
 $0.85$  & $-0.10064806619972$  &   $ 10^{-9}$\\
 \hline
 $0.9$  &  $-0.09407987936101$  &   $10^{-10} $\\
 \hline
 $0.95$  & $-0.09012580241255$  &   $10^{-12} $\\
 \hline
 $1$  &    $-0.08764080447788$  &   $10^{-13} $\\
 \hline
 $1.05$  & $-0.08601844207001$  &   $10^{-13} $\\
 \hline
 $1.1$  & $-0.084921959061503$  &   $ 10^{-13}$\\
 \hline
 $1.15$  & $-0.08415727621616$  &   $10^{-13}$\\
 \hline
 $1.21$  & $-0.08351819356943$  &   $10^{-13}$\\
 \hline
 $1.25$  & $-0.08320530895649$  &   $10^{-13}$\\
 \hline
 $1.3$  & $-0.082901918346955$  &   $10^{-13}$\\
 \hline
\end{tabular}\label{table_1}
\bigskip
\caption{Numerical data for the coupling constant dependence of the Gibbs potential around the free fermion point, when 
$\beta_1=3 \cdot 10^{-1},  {\beta}_{-1}= 10^{-1},   \beta_3=5/6, \, {\beta}_{-3}=1/6$. }
\label{text1}
\end{center}
\end{table}
\normalsize


\begin{table}[h]
\begin{center}
\begin{tabular}{|c|c|c|c|}
\hline
$ p $ & $ {\cal G}(\underline{\beta}|p) $  & $\mbox{num. error}$   \tabularnewline
 \hline
 $1$  &                 $-0.08764080447213$  &   $10^{-13} $\\
 \hline
 $1-1 \cdot 10^{-3}$  & $-0.08768059024630$  &   $10^{-13} $\\
 \hline
 $1-2 \cdot 10^{-3}$  & $-0.08772071608898$  &   $ 10^{-13}$\\
 \hline
 $1-3 \cdot 10^{-3}$  & $-0.08776118549140$  &   $10^{-13}$\\
 \hline
 $1+1 \cdot 10^{-3}$  & $-0.08760135531564$  &   $10^{-13}$\\
 \hline
 $1+2 \cdot 10^{-3}$  & $-0.08756223936590$  &   $10^{-13}$\\
 \hline
 $1+3 \cdot 10^{-3}$  & $-0.08752345325135$  &   $10^{-13}$\\
 \hline
\end{tabular}\label{table_2}
\bigskip
\caption{Numerical data for demonstrating the continuity of the Gibbs potential and its first two derivatives around the free fermion point, when 
$\beta_1=3 \cdot 10^{-1}, $   $ {\beta}_{-1}= 10^{-1}, \beta_3=5/6, \, {\beta}_{-3}=1/6$. }
\label{text2}
\end{center}
\end{table}
\normalsize


\clearpage

\newpage


\end{document}